\documentstyle[twoside,fleqn,espcrc2,psfig,graphics]{article}
\newcommand{\AmS}{{\protect\the\textfont2
  A\kern-.1667em\lower.5ex\hbox{M}\kern-.125emS}}

\title{A generalized maximal abelian gauge in SU(3) lattice gauge theory
\thanks{Poster presented by W. Tucker}
}

\input epsf

\author{William W. Tucker and John D. Stack
\address{Department of Physics, University of Illinois at 
Urbana-Champaign, 1110 West Green Street, Urbana, Illinois, 68101}
\thanks{This work was supported in part by the National Science 
Foundation. The computations on the SV1 and Regatta HPC systems
(Texas, Austin) and SP system (Michigan) were supported by NPACI.
}}
       
\begin{document}

\begin{abstract}
We introduce a generalized Maximum Abelian Gauge (MAG). We work
with this new gauge on $12^4$ lattices for $\beta=5.7,5.8$
and $16^4$ lattices for $\beta=5.9,6.0$. We also introduce a
form of abelian projection related to the generalized MAG.
We measure $U(1) \times U(1)$ wilson loops and single color
magnetic current densities.
\end{abstract}

\maketitle

Abelian dominance is the idea that the confining physics of an $SU(N)$
gauge theory can be explained by the $U(1)^{N-1}$ degrees of freedom.
In lattice gauge theory, this is tested by taking link variables with
values in the group $SU(N)$ and projecting them onto the $U(1)^{N-1}$
subgroup. In general, some sort of gauge-fixing procedure is used first.
The most commonly used gauge is the maximum abelian gauge (MAG).
Gauge-fixing to this
gauge can be thought of as maximizing the diagonal components of the
$SU(N)$ matrix, perhaps squeezing as much physics as possible into the
abelian degrees of freedom.

To date, most of the work on the MAG has been done in $SU(2)$ LGT.
Here the lattice functional that is maximized can be expressed
\begin{equation} 
G_{mag}^{su2}=\frac{1}{2N_l}\sum_{x,\mu}
{\rm Tr}\left[U^{\dagger}_{\mu}(x)\sigma_{3}U_{\mu}(x)
\sigma_{3}\right],
\label{mag2}
\end{equation}
where $N_l$ is the number of lattice links.
After sweeping over the lattice performing gauge transformations many
times, the gauge-fixed $SU(2)$ link variables are projected onto the
$U(1)$ subgroup and calculations of observables are carried out as in
$U(1)$ gauge theory. The projected configurations typically give values
of the string tension that are about 10\% higher than the full $SU(2)$
value at one gauge copy per configuration. These values come down the
$SU(2)$ value when Gribov effects are taken into account.

The MAG in $SU(3)$ is constructed similarly to $SU(2)$. We expect the
physics to be similar for both of these theories, but the $SU(3)$
structure is more complicated. The simplest MAG functional may now
be expressed as
\begin{eqnarray}
G_{mag}^{su3}& = & \frac{1}{4N_l}\left\{
\sum_{x,\mu}
{{\rm Tr}\left[U^{\dagger}_{\mu}(x)\lambda_{3}U_{\mu}(x)
\lambda_{3}\right]}+\right. \nonumber \\
& &\left.\sum_{x,\mu}{{\rm Tr}\left[U^{\dagger}_{\mu}(x)\lambda_{8}U_{\mu}(x)
\lambda_{8}\right]} \right\}.
\label{mag3}
\end{eqnarray}
It has been found that this form of the MAG in $SU(3)$ gives string
tensions about 10\% lower than the full $SU(2)$ value. This
discrepancy increases when Gribov effects are taken into account
\cite{jrwsu3}.

We begin our construction of a new MAG functional by noting that
the $SU(2)$ MAG functional can be expressed as the gauge field
coupled to an adjoint Higgs field rotated to the $\sigma_3$
direction. An equivalent $SU(3)$ functional would have the Higgs
field rotated so as to take values in the Cartan subalgebra of
$SU(3)$. The functional from Eq.(\ref{mag3}) can not be
duplicated with a single Higgs field, but can be recreated using
two or more Higgs fields.

\section{Generalized Maximal Abelian Gauge}

On the lattice, our new functional will take the form
\begin{eqnarray}
G_{higgs}^{su3} = \frac{1}{N_{h}N_{l}}\sum_{i,x,\mu}
{{\rm Tr}\left[U^{\dagger}_{\mu}(x)\Phi_{i}U_{\mu}(x)
\Phi_{i}\right]}
\label{maghiggs}
\end{eqnarray}
where the Higgs fields are
\begin{equation}
\Phi_{i} = \frac{1}{\sqrt{2}}\left[ \lambda_{3}{\rm cos}\chi_{i} + 
\lambda_{8}{\rm sin}\chi_{i} \right],
\label{magphi}
\end{equation}
and $N_h$ is the number of Higgs fields. In this equation, we
could allow the value of $\chi_i$ to be site-dependent
\cite{slovakia}.
Here, we will deal with the case of constant Higgs fields.
If we have two fixed Higgs field, with \mbox{$\chi_{1}=0^o$} and
\mbox{$\chi_{2}=90^o$} we get back the simple MAG with functional
given by Eq.(\ref{mag3}).

The gauge-fixing procedure consists of maximizing this functional
on a site-by-site basis, one $SU(2)$ subgroup at a time. This can
be done analytically. Overrelaxation was used to speed up
convergence \cite{mandula}. For fixed $\chi$, an overrelaxation
parameter $\omega=1.85$ was used.

\section{Abelian Projection}

The act of projection is another instance in which the $SU(3)$ case
has substantial complications that do not arise in $SU(2)$. In
$SU(2)$, the diagonal elements are complex conjugates of one
another. Taking the phase and using it as a $U(1)$ variable is a
straightforward process.

For $SU(3)$ case, simply taking the phases of the diagonal entries
does not give a $U(1) \times U(1)$ matrix. Here,  we have used an
``optimal'' method \cite{berlin}. For the simple MAG of
\mbox{Eq. (\ref{mag3}),} this method finds $U'$,
the element of $U(1) \times U(1)$ that maximizes the quantity 
\begin{equation}
{\rm Re}\left({\rm Tr}\left[U'D_{\mu}^{\dag}(x) \right] \right),
\label{optproject}
\end{equation}
where $D_{\mu}(x)$ is the diagonal component of the original $SU(3)$
matrix.

The generalized MAG suggests a generalized form of projection that
takes into account the Higgs field we used for gauge-fixing. We
now find $U'$ that maximizes the quanitity
\begin{equation}
\sum_i{{\rm Re}\left({\rm Tr}\left[\Phi_{i}(x)U'
\Phi_{i}(x+\hat{\mu})D_{\mu}^{\dag}(x)
\right] \right)}.
\label{genproject}
\end{equation}
We note that for the case of the simple MAG this is the
same as the optimal method used previously. As long as the
Higgs fields are constant this just amounts to a
rescaling of the of the components of $D_{\mu}(x)$. It also
demonstrates the importance of the method of projection used
in the variable Higgs case.

For the case of constant Higgs fields, it is useful to
talk of the functional
\begin{equation}
P_{\phi}=\frac{1}{N_{h}N_{l}}\sum_{i,x,\mu}
{{\rm Re}\left({\rm Tr}\left[\Phi_{i}^{2}U'(x)D_{\mu}^{\dag}(x)
\right] \right)}.
\label{genproject2}
\end{equation}
as an indication of how far we have to project to get to
the abelian submanifold.

\section{Magnetic Currents}

Magnetic currents are extracted for each $SU(3)$ color by applying the
Toussaint DeGrand procedure to the $U(1) \times U(1)$ links.  This 
produces three magnetic currents. After extracting the 
magnetic currents, monopole Wilson loops are calculated for each color.  This
part of the calculation proceeds exactly as in $U(1)$ or $SU(2)$ lattice
gauge theory \cite{js-rw,stack}.

\begin{table*}[htb]
% space before first and after last column: 1.5pc
% space between columns: 3.0pc (twice the above)
\setlength{\tabcolsep}{.2pc}
% -----------------------------------------------------
% adapted from TeX book, p. 241
\catcode`?=\active \def?{\kern\digitwidth}
% -----------------------------------------------------
%\caption{String tensions for SU(3) MAG}
\caption{$12^4$ results for $\beta=5.7$, $\beta=5.8$.}
\label{tab1}
\begin{tabular*}{\textwidth}{@{}l@{\extracolsep{\fill}}ccccccc|ccccccc}
\hline
%                 & \multicolumn{2}{l}{Pilot plant} 
%                 & \multicolumn{2}{l}{Full scale plant} \\
%\cline{2-3} \cline{4-5}
                 & \multicolumn{7}{c|}{$\beta=5.7$}
                 & \multicolumn{7}{c}{$\beta=5.8$} \\
                 & \multicolumn{1}{c}{$\chi_{1}$}
                 & \multicolumn{1}{c}{$\chi_{2}$}
                 & \multicolumn{1}{c}{$P_{\phi}$}
                 & \multicolumn{1}{c}{$C(3,2)$}
                 & \multicolumn{1}{c}{$f_1$}
                 & \multicolumn{1}{c}{$f_2$}
                 & \multicolumn{1}{c|}{$f_3$}
                 & \multicolumn{1}{c}{$\chi_{1}$}
                 & \multicolumn{1}{c}{$\chi_{2}$}
                 & \multicolumn{1}{c}{$P_{\phi}$}
                 & \multicolumn{1}{c}{$C(3,2)$}
                 & \multicolumn{1}{c}{$f_1$}
                 & \multicolumn{1}{c}{$f_2$}
                 & \multicolumn{1}{c}{$f_3$}  \\
\hline
&  $0^o$  & $90^o$ & 0.885 & 0.169(13) & 0.040 & 0.040 & 0.040
&  $0^o$  & $90^o$ & 0.892 & 0.118(11) & 0.023 & 0.023 & 0.023 \\
&  $0^o$  & $60^o$ & 0.886 & 0.173(13) & 0.038 & 0.043 & 0.043
&  $0^o$  & $60^o$ & 0.892 & 0.120(12) & 0.022 & 0.025 & 0.025 \\
&  $15^o$ & $45^o$ & 0.886 & 0.179(15) & 0.040 & 0.047 & 0.047
&  $15^o$ & $45^o$ & 0.893 & 0.123(13) & 0.023 & 0.027 & 0.027 \\
& $-15^o$ & $15^o$ & 0.883 & 0.189(17) & 0.047 & 0.047 & 0.081 
& $-15^o$ & $15^o$ & 0.890 & 0.136(11) & 0.029 & 0.030 & 0.053 \\
&  $15^o$ & $15^o$ & 0.872 & 0.181(15) & 0.043 & 0.101 & 0.134
&  $15^o$ & $15^o$ & 0.879 & 0.133(12) & 0.026 & 0.080 & 0.102 \\
&   $0^o$ &  $0^o$ & 0.881 & 0.203(13) & 0.055 & 0.055 & 0.099
&   $0^o$ &  $0^o$ & 0.888 & 0.147(13) & 0.036 & 0.036 & 0.067 \\
\hline

%\multicolumn{5}{@{}p{120mm}}{Reprinted from: G.M. Ritcey,
%                             Tailings Management,
%                             Elsevier, Amsterdam, 1989, p. 635.}
\end{tabular*}
\end{table*}

\begin{table*}[htb]
% space before first and after last column: 1.5pc
% space between columns: 3.0pc (twice the above)
\setlength{\tabcolsep}{.2pc}
% -----------------------------------------------------
% adapted from TeX book, p. 241
\catcode`?=\active \def?{\kern\digitwidth}
% -----------------------------------------------------
%\caption{String tensions for SU(3) MAG}
\caption{$16^4$ results for $\beta=5.9$, $\beta=6.0$.}
\label{tab2}
\begin{tabular*}{\textwidth}{@{}l@{\extracolsep{\fill}}ccccccc|ccccccc}
\hline
%                 & \multicolumn{2}{l}{Pilot plant} 
%                 & \multicolumn{2}{l}{Full scale plant} \\
%\cline{2-3} \cline{4-5}
                 & \multicolumn{7}{c|}{$\beta=5.9$}
                 & \multicolumn{7}{c}{$\beta=6.0$} \\
                 & \multicolumn{1}{c}{$\chi_{1}$}
                 & \multicolumn{1}{c}{$\chi_{2}$}
                 & \multicolumn{1}{c}{$P_{\phi}$}
                 & \multicolumn{1}{c}{$C(3,2)$}
                 & \multicolumn{1}{c}{$f_1$}
                 & \multicolumn{1}{c}{$f_2$}
                 & \multicolumn{1}{c|}{$f_3$}
                 & \multicolumn{1}{c}{$\chi_{1}$}
                 & \multicolumn{1}{c}{$\chi_{2}$}
                 & \multicolumn{1}{c}{$P_{\phi}$}
                 & \multicolumn{1}{c}{$C(3,2)$}
                 & \multicolumn{1}{c}{$f_1$}
                 & \multicolumn{1}{c}{$f_2$}
                 & \multicolumn{1}{c}{$f_3$}  \\
\hline
&  $0^o$  & $90^o$ & 0.898 & 0.084(07) & 0.012 & 0.012 & 0.012
&  $0^o$  & $90^o$ & 0.903 & 0.066(06) & 0.008 & 0.007 & 0.007 \\
&  $0^o$  & $60^o$ & 0.898 & 0.087(07) & 0.012 & 0.014 & 0.015
&  $0^o$  & $60^o$ & 0.903 & 0.067(06) & 0.007 & 0.008 & 0.008 \\
&  $15^o$ & $45^o$ & 0.898 & 0.087(07) & 0.013 & 0.016 & 0.015
&  $15^o$ & $45^o$ & 0.903 & 0.069(06) & 0.008 & 0.009 & 0.009 \\
& $-15^o$ & $15^o$ & 0.896 & 0.101(10) & 0.018 & 0.018 & 0.033 
& $-15^o$ & $15^o$ & 0.901 & 0.079(08) & 0.011 & 0.011 & 0.021 \\
&  $15^o$ & $15^o$ & 0.885 & 0.105(08) & 0.016 & 0.064 & 0.079
&  $15^o$ & $15^o$ & 0.890 & 0.088(08) & 0.010 & 0.052 & 0.062 \\
&   $0^o$ &  $0^o$ & 0.894 & 0.109(10) & 0.023 & 0.023 & 0.045
&   $0^o$ &  $0^o$ & 0.899 & 0.086(10) & 0.015 & 0.016 & 0.030 \\
\hline

%\multicolumn{5}{@{}p{120mm}}{Reprinted from: G.M. Ritcey,
%                             Tailings Management,
%                             Elsevier, Amsterdam, 1989, p. 635.}
\end{tabular*}
\end{table*}

\section{Results}

We gauge-fixed 40 configurations on $12^4$ lattices for each of
$\beta=5.7,5.8$. We also gauge-fixed 20 configurations on
$16^4$ lattices with $\beta=5.9,6.0$. 
A stopping condition similar to that used in \cite{jrwsu3}.
The value of the projection functional $P_{\phi}$, the
$U(1) \times U(1)$ 3-2 Creutz ratio, and the
fractions of links with single color magnetic current are given
in Tables 1 and 2. The Creutz ratios may be compared to the full
$SU(3)$ string tensions 0.168(1), 0.109(1), 0.073(1), 0.054(1)
for \mbox{$\beta=5.7,5.8,5.9,6.0$} respectively \cite{bornetal}.

Our principle observation is that with two fixed Higgs fields, there
are two broad classes of gauges that result. MAG-like gauges result
when the Higgs fields are separated by an unbroken SU(2) symmetry,
that is by a $\lambda_8$-like Higgs field. An example of an MAG-like
gauge would be \mbox{$\chi_1=15^o$,} \mbox{$\chi_2=45^o$.} MAG-like gauges are
characterized by a high value for the functional $P_{\phi}$
and roughly equal color distribution for magnetic currents. 
Higgs-like gauges, so named because they include the case of a single
Higgs field, occur when the \mbox{$\chi$'s} are not so separated. An
example of such a gauge would be \mbox{$\chi_1=-15^o$,}
\mbox{$\chi_2=15^o$.} They are characterized by a range of lower
$P_{\phi}$ values and and an asymmetric distribution of magnetic current.
It should be noted that values for the string tension as determined by
Creutz ratios from those Wilson Loops were also unequal.

\section{Conclusion}

The MAG-type gauges proved to be better than
the Higgs-type gauges, at least in terms of how much information
was maintained in the projection. All the MAG-type gauges
similar Creutz ratios. One note is that there were MAG-type gauges
that had slightly higher $P_{\phi}$ than the original MAG. These
gauges also had slightly higher Creutz ratios.

We expect to gather better statistics on all of these cases.
There is also a lot to be learned about the variable Higgs
case.

% \vspace{-0.3cm}

%\begin{figure}[htb]
%\psfig{file=figure1.eps,height=8.cm}
%\vspace{-1.0cm}
%\caption{V(R) vs R for MAG ($U(1) \times U(1)$ and monopoles)
%and ICG \mbox{(P-vortices)} for $SU(3)$}
%\label{fig1}
%\end{figure}

%\vspace{-0.4cm}

\end{document}